\documentstyle[preprint,aps]{revtex}
\tightenlines
\begin{document}
\preprint{\today}
\title{Testing Mode-Coupling Theory for a Supercooled Binary
Lennard-Jones Mixture I: The van Hove Correlation Function}
\author{Walter Kob\cite{wkob}}
\address{Institut f\"ur Physik, Johannes Gutenberg-Universit\"at,
Staudinger Weg 7, D-55099 Mainz, Germany}
\author{Hans C. Andersen\cite{hca}}
\address{Department of Chemistry, Stanford University, Stanford,
California 94305}
\maketitle

\begin{abstract}
We report the results of a large scale computer simulation of a binary
supercooled Lennard-Jones liquid. We find that at low temperatures the
curves for the mean squared displacement of a tagged particle for
different temperatures fall onto a master curve when they are plotted
versus rescaled time $tD(T)$, where $D(T)$ is the diffusion constant.
The time range for which these curves follow the master curve is
identified with the $\alpha$-relaxation regime of mode-coupling theory
(MCT). This master curve is fitted well by a functional form suggested
by MCT. In accordance with idealized MCT, $D(T)$ shows a power-law
behavior at low temperatures. The critical temperature of this
power-law is the same for both types of particle and also the critical
exponents are very similar.  However, contrary to a prediction of MCT,
these exponents are not equal to the ones determined previously for the
divergence of the relaxation times of the intermediate scattering
function [Phys. Rev. Lett. {\bf 73}, 1376 (1994)].  At low temperatures
the van Hove correlation function (self as well as distinct part) shows
hardly any sign of relaxation in a time interval that extends over
about three decades in time.  This time interval can be interpreted as
the $\beta$-relaxation regime of MCT. From the investigation of these
correlation functions we conclude that hopping processes are not
important on the time scale of the $\beta$-relaxation for this system
and for the temperature range investigated.  We test whether the
factorization property predicted by MCT holds and find that this is
indeed the case for all correlation functions investigated. The distance
dependence of the critical amplitudes are in qualitative accordance
with the ones predicted by MCT for some other mixtures. The
non-gaussian parameter for the self part of the van Hove correlation
function for different temperatures follows a master curve when plotted
against time $t$.

\end{abstract}

\pacs{PACS numbers: 61.20.Lc, 61.20.Ja, 64.70.Pf, 51.10.+y}
%
%
\section{Introduction}
\label{sec:I}
About a decade has passed since two of the most seminal papers in
the recent history of the field of the glass transition and supercooled
liquids were published. In these papers, one by Bengtzelius, G\"otze,
and Sj\"olander and the other by Leutheusser, it was proposed that the
glass transition could be understood as the singular behavior of the
solution of the equations of motion of the dynamic structure factor,
the so-called mode-coupling equations\cite{bgs84lh84}.  These equations
are the simplified versions of certain nonlinear equations of motion
that were derived in the seventies in order to describe the dynamics of
simple liquids at high densities\cite{kinetictheodensliquids}.
Subsequently G\"otze, Sj\"ogren and many others analyzed these
mode-coupling equations in order to work out the details of the
singular behavior of their solutions
\cite{mcttheory,barratwgal89,fuchsdiss93,fuchs92}. Today the sum of all
these results is known as mode-coupling theory (MCT), and a review of
them can be found in some recent review
articles\cite{bibles,schilling94}. The theory has stimulated a
remarkable amount of experimental and computer simulation work,
with various groups looking
for the signs of this singularity
in many different kinds of
systems\cite{mctexp,roux89,signorinijlbmlk90,lewis91,wahnstrom91,baschnagel94,kob94a,kob94b,brakkee90,cummins93,du94,baschfu94}.
The result of all these experiments and simulations is that the theory
appears to be able to rationalize the dynamical behavior of some glass
forming materials in an amazingly convincing way. However, the dynamical
behavior of other glass forming  materials seems to be described by MCT only
poorly.  So far it is still not clear yet for what kind of system the
prediction of the theory hold and for which systems they do not. Even
if for a particular system some of the predictions of the theory hold,
it is not certain that also the other predictions made by MCT will
hold. Therefore it is clear that still much has to be learned about the
applicability of this theory. However, it seems that there is agreement
on at least one point, namely that the dynamical singularity predicted
by MCT is not the same as the {\em laboratory} glass transition. The
latter occurs at a temperature $T_{g}$ that is defined as the
temperature at which the viscosity of the material is $10^{13}$ Poise.
Below this temperature the material can no longer come to thermodynamic
equilibrium because its relaxation times are longer than the time scale
of typical experiments.  However, it is found that if experiments see a
singular behavior that can be interpreted as the singularity predicted
by MCT, then the temperature at which this singular behavior is
observed is about 30-50K above $T_{g}$. Furthermore it has also been
found empirically that the predictions of MCT seems to work better for
fragile glass formers than for strong glass formers.  For fragile glass
formers, a plot of the logarithm of the viscosity vs. $1/T$ is curved,
and the temperature of the MCT singularity is often near the
temperature at which the plot shows a pronounced bend.  At this
temperature the viscosity changes its behavior from a weak dependence
on temperature to a strong dependence on temperature as the temperature
is lowered.  Thus it can be that this bend in the viscosity is a
signature of the singularity predicted by MCT. This bend occurs at
viscosities which are about 1-100 Poise or relaxation times that are
around $10^{-11}-10^{-9}$ seconds.  Thus the singularity predicted by
MCT is not the laboratory glass transition but rather is an anomalous
dynamic behavior in supercooled liquids for fragile glass forming
materials that takes place at temperatures above the glass transition
temperature.\par

Computer simulations are particularly well suited to test the
predictions of MCT, since they allow access to the full information on
the system at any time of the simulation. This in turn permits the
calculation of many different types of correlation functions, some of
which are not experimentally measurable but about which the theory
makes definite predictions, and thus very stringent tests of the theory
become possible. Furthermore the measurement of these correlation
functions is very straightforward in that they are computed directly
from the positions and velocities of the particles. Thus no
theoretical model or assumption, as has to be used to explain, e.g.,
the scattering mechanism in light scattering experiments, is needed.\par

A severe drawback of most computer simulations is the limited range of
times over which simulations can be performed.  This in effect makes
the cooling rates in simulations much larger than those in laboratory
experiments. As a result the simulated material falls out of
equilibrium at a higher temperature than would the corresponding real
material in the laboratory\cite{angell80}.  In other words, the glass
transition temperature in the simulation $T_{g-sim}$ is higher than the
laboratory glass transition $T_g$.  If $T_{g-sim}$ is too high, the
lack of equilibration in the simulation can obscure not only the MCT
singularity at $T_c$ but also the higher temperature signatures of the
onset of the singularity.  Thus, it is important in simulations to have
a range of temperatures, extending down as close as possible to $T_c$,
at which the system can be thoroughly equilibrated and the slow
dynamics studied.  This requires a significant amount of computation to
achieve. In the present paper we will present data for thoroughly
equilibrated systems in such a temperature range.\par

This paper presents the results of a molecular dynamics computer
simulation in order to make careful tests of whether the predictions of
MCT hold for the system under investigation.  In two previous papers we
investigated for the same system the scaling behavior of the
intermediate scattering function in the $\beta$-relaxation
regime\cite{kob94a,kob94b}.  We found that this correlation function shows
many of the features predicted by the theory. In the present paper we
will focus on the behavior of the van Hove correlation function and
test whether the predictions of MCT hold for these kind of correlation
functions as well. In a following paper\cite{kob94d} we will
investigate in detail the time and wave-vector dependence of the
intermediate scattering function in the $\alpha$-and $\beta$-relaxation
regime (defined in the next section) and also the frequency dependence
of the dynamic susceptibility.  Thus the sum of the results of these
investigations will allow us to make a stringent test on whether or not
MCT is able to rationalize the dynamical behavior of the system
studied.\par

The present paper is organized as follows:  In Sec. II we will
summarize those predictions of MCT that are relevant to understand the
results of this work. In Sec.  III we introduce our model and give some
details on the computation.  In Sec. IV we present our results and in
Sec. V we summarize and discuss these results.

\section{Mode Coupling Theory}
In this section we give a short summary of those predictions of MCT
that are relevant for the interpretation of the results presented in
this paper. An extensive review of the theory can be found in
references \cite{bibles,schilling94}.\par

In its simplest version, also called the idealized version, mode-coupling
theory predicts the existence of a critical temperature $T_{c}$ above
which the system shows ergodic behavior and below which the system is
no longer ergodic. All the predictions of the theory which are
considered in this paper are of an asymptotic nature in the sense that
they are valid only in the vicinity of $T_{c}$. For temperatures close
to $T_{c}$, MCT makes precise predictions about the dynamical behavior
of time correlation functions $\phi(t)=\langle X(0)Y(t)\rangle$ between
those dynamical variables $X$ and $Y$ that have a nonzero overlap
with $\delta \rho ({\mbox{\boldmath $q$}})$, the fluctuations of the
Fourier component of the density for wave vector {\boldmath $q$}, i.e.
for which $\langle \delta \rho ({\mbox{\boldmath $q$}}) X \rangle$ and
$\langle \delta \rho ({\mbox{\boldmath $q$}}) Y \rangle$ is
nonvanishing. Here $\langle \: \rangle$ stands for the canonical
average. In particular the theory predicts that for $T>T_{c}\,$
$\phi(t)$ should show a two step relaxation behavior, i.e.  the
correlation function plotted as a function of the logarithm of time
should show a decay to a plateau value, for intermediate times,
followed by a decay to zero at longer times.  The time interval in
which the correlation functions are close to this plateau is called the
$\beta$-relaxation region. Despite a similar name this region should
not be confused with the $\beta$-relaxation process as described by
Johari and Goldstein\cite{joharietal}.\par

Furthermore the theory predicts that in the vicinity of the plateau the
so-called ``factorization-property'' holds. This means that the correlator
$\phi(t)$ can be written in the form
\begin{equation}
\phi(t)=f^{c}+h G(t)\quad ,
\label{eq1}
\end{equation}
where $f^{c}$ is the height of the plateau at $T_{c}$, $h$ is some
amplitude that depends on the correlator but not on time and the
function $G(t)$ depends on time and temperature {\em but not} on
$\phi$. Thus for a given system $G(t)$ is a universal function for all
correlators satisfying the above mentioned condition. The details of
the function $G(t)$ depend on a system specific parameter $\lambda$,
the so-called exponent parameter.  In principle $\lambda$ can be
calculated if the structure factor of the system is known with
sufficient precision, but since this is rarely the case for real
experiments, it is in most cases treated as a fitting parameter.  For
all values of $\lambda$ the theory predicts that for certain time
regions the functional form of $G(t)$ is well approximated by two
power-laws. In particular it is found that for those times for which
the correlator is still close to the plateau but has started to deviate
from it, $G(t)$ is given by the so-called von Schweidler law:
\begin{equation}
G(t)=-B(t/\tau)^{b}\quad.
\label{eq2}
\end{equation}
where $B$ is a constant that can be computed from $\lambda$. The
relaxation time $\tau$ is the relaxation time of the so-called
$\alpha$-relaxation, i.e. the relaxation at very long times where the
correlator decays to zero. The exponent $b$, often called the von
Schweidler exponent, can be computed if the value of $\lambda$ is known
and is therefore {\it not} an additional fitting parameter.\par

In the $\alpha$-relaxation regime MCT predicts that the correlation
functions obey the time temperature superposition principle, i.e.
\begin{equation}
\phi(t)=F(t/\tau)\quad,
\label{eq2n}
\end{equation}
where the overwhelming part of the temperature dependence of the right
hand side is given by the temperature dependence of $\tau$.
Equation~(\ref{eq2n}) says that if the correlation function for
different temperatures are plotted versus $t/\tau(T)$ they will fall
onto the master curve $F(t/\tau)$. Note that the time range in which
the von Schweidler law is observed belongs to the late
$\beta$-relaxation regime as well as to the early $\alpha$-relaxation
regime.\par

In addition MCT predicts that the diffusion constant $D$ shows a
power-law behavior as a function of temperature with the critical
temperature $T_{c}$:
\begin{equation}
D\propto (T-T_{c})^{\gamma}\quad,
\label{eq3}
\end{equation}
where $\gamma$ can also be computed once $\lambda$ is known.\par

Note that some of these predictions of MCT are valid only for the
simplest (or idealized) version of the theory in which the so-called
hopping processes are neglected. If these processes are present some of
the statements made above have to be modified. However, below we will
give evidence that for the system under investigation hopping processes
are not important in the temperature range investigated and that
therefore the idealized version of the theory should be applicable.\par

Furthermore it has to be emphasized that MCT assumes that the system
under investigation is in {\em equilibrium}. Thus great care should be
taken to equilibrate the system properly. A recent computer
simulation of a supercooled polymer system has shown that
nonequilibrium effects can completely change the behavior of the
correlation functions\cite{baschnagel94}. Thus a comparison of the
predictions of MCT with the results of a simulation in which
nonequilibrium effects are still present becomes doubtful at best.

\section{Model and Computational Procedures}
\label{sec:III}
In this section we introduce the model we investigated and give some
of the details of the molecular dynamics simulation.

The system we are studying in this work is a binary mixture of
classical particles. Both types of particles (A and B) have the same
mass $m$ and all particles interact by means of a Lennard-Jones
potential, i.e. $V_{\alpha\beta}(r)=4\epsilon_{\alpha\beta}
[(\sigma_{\alpha\beta}/r)^{12}-(\sigma_{\alpha\beta} /r)^{6}]$ with
$\alpha,\beta \in\{A,B\}$.  The reason for our choice of a mixture was
to prevent the crystallization of the system at low temperatures.
However, as we found out in the course of our work, choosing a binary
mixture is by no means sufficient to prevent crystallization, if the
system is cooled slowly. In particular we found that a model that has
previously been used to investigate the glass transition
\cite{ernst91}, namely a mixture of 80\% A particles and 20\% B
particles with $\epsilon_{AA}=\epsilon_{AB}=\epsilon_{BB}$, and
$\sigma_{BB}=0.8\sigma_{AA}$, and $\sigma_{AB}=0.9\sigma_{AA}$,
crystallizes at low temperatures, as evidenced by a sudden drop in the
pressure.  In order to obtain a model system that is less prone to
crystallization, we adjusted the parameters in the Lennard-Jones
potential in such a way that the resulting potential is similar to one
that was proposed by Weber and Stillinger to describe amorphous
Ni$_{80}$P$_{20}$\cite{stillweber86}. Thus we chose
$\epsilon_{AA}=1.0$, $\sigma_{AA}=1.0$, $\epsilon_{AB}=1.5$,
$\sigma_{AB}=0.8$, $\epsilon_{BB}=0.5$, and $\sigma_{BB}=0.88$. The
number of particles of type A and B were 800 and 200, respectively.
The length of the cubic box was 9.4 $\sigma_{AA}$ and periodic boundary
conditions were applied. In order to lower the computational burden we
truncated and shifted the potential at a cut-off distance of
$2.5\sigma_{\alpha\beta}$. In the following all the results will be
given in reduced units, i.e. length in units of $\sigma_{AA}$, energy
in units of $\epsilon_{AA}$ and time in units of
$(m\sigma_{AA}^2/48\epsilon_{AA})^{1/2}$. For Argon these units
correspond to a length of 3.4\AA, an energy of 120K$k_B$ and a time of
$3\cdot10^{-13}$s.\par

The molecular dynamics simulation was performed by integrating the
equations of motion using the velocity form of the Verlet algorithm
with a step size 0.01 and 0.02 at high ($T\geq 1.0$) and low ($T\leq
0.8$) temperatures, respectively. These step sizes were sufficiently
small to reduce the fluctuation of the total energy to a negligible
fraction of $k_{B}T$. The system was equilibrated at high temperature
($T=5.0$) where the relaxation times are short. Changing the
temperature of the system to a temperature $T_{f}$ was performed by
coupling it to a stochastic heat bath, i.e.  every 50 steps the
velocities of the particles were replaced with velocities that were
drawn from a Boltzmann distribution corresponding to the temperature
$T_{f}$. This was done for a time period of length $t_{equi}$, which
was chosen to be larger than the relaxation time of the system at the
temperature $T_{f}$. After this change of temperature we let the system
propagate with constant energy, i.e.  without the heat bath, for a time
that was also equal to $t_{equi}$, in order to see that there was no
drift in temperature, pressure, or potential energy. If no drift was
observed, we considered the final state to correspond to an equilibrium
state of the system at the temperature $T_{f}$, and we used this final
state as the initial state for a molecular dynamics trajectory.  In
this trajectory, there was no coupling to a heat bath; it was a
constant energy trajectory, and the results were used to provide the
correlation function data discussed below.  The temperatures we studied
were $T=5.0$, 4.0, 3.0, 2.0, 1.0, 0.8, 0.6, 0.55, 0.50, 0.475, and
0.466. At the lowest temperatures the length of the run was $10^{5}$
time units. Thus, again assuming Argon units, the present data covers
the time range from $3\cdot 10^{-15}$s to $3\cdot10^{-8}$s. At the
lowest temperature the equilibration time $t_{equi}$ was
$0.6\cdot10^{5}$ time units. Thus if we define the cooling rate to be
the difference of the starting temperature and the final temperature
divided by the time for the quench, the smallest cooling rate (used to
go from the second lowest temperature to the lowest one) is $1.5\cdot
10^{-7}$. In the case of Argon this smallest cooling rate would
correspond to $6\cdot 10^{7}$K/s. Thus, although this cooling rate is
still very fast it is smaller than the fastest cooling rate achievable
in experiments and about an order of magnitude smaller that the one
used in previous computer simulations.\par

In order to improve the statistics, we performed eight different runs
at each temperature, each of which was equilibrated separately in the above
described way, and averaged the results.  Each of these runs originated from
a different point in configuration space.  The thermal history of these
starting points differed significantly from one to another. In
particular this history sometimes included periods in which we reheated
the system after having cooled it to low temperatures and had it
equilibrated at these temperatures or, in some other cases, included
periods in which we cooled it with twice the normal cooling rate.
Despite these different thermal histories the results we obtained from
these eight different runs were the same to within statistical
fluctuations. Thus we have good evidence that the results reported in
this work are all {\em equilibrium} properties of the system and are
not dependent on the way we prepared the system at a given
temperature.\par

Most of the results presented in this paper deal with the van Hove
correlation function (self and distinct part), which are defined
below.  This space-time correlation function is the one that is most
easily obtained from molecular dynamics computer simulations. As
mentioned in the previous paragraph we averaged our results over at
least eight different runs.  Since the resulting correlation functions
$\phi(r,t)$ still showed some short wavelength noise in $r$ even after
we did this averaging, we smoothed the data {\em in space} by means of a
spline under tension\cite{reinsch67}. No smoothing was done in time,
since the data was so smooth that such a treatment seemed not
necessary.

\section{Results}
\label{sec:IV}
In this section we present the results of our simulation. In the
first part we will deal with time independent quantities and in the
second part with time dependent quantities.

In many computer simulations dealing with the glass transition it is
observed that there exists a temperature $T_{g-sim}$ at which certain
thermodynamic quantities, like e.g. the potential energy per particle,
show some sort of discontinuity (e.g. a rapid change of slope) when
plotted versus temperature (see, e.g., \cite{brakkee90}). The
temperature where these features are observed have nothing to do with
the laboratory glass transition temperature $T_{g}$, the temperature
where the viscosity of the material has a value of $10^{13}$Poise,
since $T_{g-sim}$ is usually significantly higher than $T_{g}$. The
physical significance of $T_{g-sim}$ is that at this temperature the
system under investigation has fallen out of equilibrium, or in other
words has undergone a glass transition, on the time scale of the
computer simulation.  Obviously $T_{g-sim}$ depends on the cooling rate
with which the system was cooled and on the thermal history of the
system.\par

In Fig.~\ref{fig1} we show the pressure, the total energy, and the
potential energy of our system as a function of temperature. In order
to expand the temperature scale at low temperatures, we plot these
quantities versus $T^{-1}$. We recognize from this figure that there is
no temperature at which one of these quantities shows any sign of an
anomalous behavior. Thus we can conclude that in this simulation the
system is not undergoing a glass transition, i.e. that for the
effective cooling rates used in this work, $T_{g-sim}$ is less than the
lowest temperature investigated. Thus this is evidence that we are able
to equilibrate the system even at the lowest temperatures. Stronger
evidence for equilibration will be presented below.

One of the simplest time dependent quantities to measure in a molecular
dynamics simulation is the mean squared displacement (MSD) $\langle
r^{2}(t)\rangle$ of a tagged particle, i.e. $\langle r^{2}(t)\rangle =
\langle |\mbox{\boldmath $r$}(t)-\mbox{\boldmath $r$}(0)|^{2} \rangle$.
In Fig.~\ref{fig2} we show this quantity for the particles of type A
versus time in a double logarithmic plot. The corresponding plot for
the B particles is very similar. The curves to the left correspond to
high temperatures and those to the right to low temperatures. We
recognize that for short times all the curves show a power-law behavior
with an exponent of two . Thus this is the ballistic motion of the
particles. At high temperatures this ballistic motion goes over
immediately into a diffusive behavior (power-law with exponent one). For
low temperatures these two regimes are separated by a time regime where
the motion of the particles seems to be almost frozen in that the MSD
is almost constant and thus shows a plateau. At the lowest temperature
this regime extends from about one time unit to about $10^{3}$ time
units. Only for much longer time (note the logarithmic time scale!) the
curves show a power-law again, this time with unit slope indicating
again that the particles have a diffusive behavior on this time scale.
The fact that the length of our simulation is long enough in order to
see this diffusive behavior even at the lowest temperature is a further
indication that we are able to equilibrate the system at all
temperatures. The reader should note that for the discussion of these
different time regimes it is most helpful to plot the curves of the
MSD with a {\em logarithmic} time axis. Only in this way it is
possible to recognize that the dynamics of the system is very
different on the various time scales.\par

Note, that the value of the MSD in the vicinity of the plateau is about
0.04, thus corresponding to a distance of about 0.2. We therefore
recognize that on this time scale the tagged particle has moved only
over a distance that is significantly shorter than the next nearest
neighbor distance (which is close to one, see below). Thus it is still
trapped in the cage of particles that surrounded it at time zero, and
it takes the particle a long time to get out of this cage.  The initial
stages of this slow breakup of the cage is exactly the type of process
MCT predicts to happen during the $\beta$-relaxation. (We will later
elaborate more on this point in the discussion of the self part of the van
Hove correlation function.) Thus we can identify the time range where
we observe the plateau in the MSD with the $\beta$-relaxation regime of
MCT.\par

{}From the MSD it is now easy to compute the self diffusion constant
$D(T)$ of the particles.  (Using a plot such as Fig. 2, a straight
line, with unit slope, fit to the long time behavior of the data
intersects a vertical line at $\log_{10}t=0$ at a height of
$\log_{10}6D$.)  Since MCT predicts that diffusion constants should
have a power-law dependence on temperature at low temperatures (see
Eq.~(\ref{eq3})), we tried to make a three parameter fit with such a
functional form.  In Fig.~\ref{fig3} we show the result of this fit by
plotting $D$ versus $T-T_{c}$ in a double logarithmic way. We clearly
observe, that, in accordance with MCT, for temperatures $T\leq 1.0$ the
diffusion constants follow a power-law behavior.  The value of $T_{c}$
is 0.435, independent of the type of particle.  This independence of
$T_{c}$ of the type of particles is in accordance with the prediction
of MCT. From the value of $T_{c}$ we now can compute the small
parameter of the theory, i.e.  $\epsilon=|T-T_{c}|/T_{c}$. At the
lowest temperature $\epsilon$ is 0.07, thus quite small and therefore
it is not unreasonable to assume that we are already in the temperature
range where the asymptotic results of the theory hold. At $T=1.0$ the
value of $\epsilon$ is 1.3, which seems rather too large for the
asymptotic expansion to apply. However, it has been found in
experiments that for some systems the predictions of MCT hold for
values of $\epsilon$ of at least 0.5\cite{du94}.  Therefore our finding
is not that astonishing.  Also, by investigating the relaxation time of
the intermediate scattering function we found that the asymptotic
behavior at low temperatures is obtained for this quantity only for
$T\leq 0.6$\cite{kob94a,kob94d}. This corresponds to a value of
$\epsilon$ of 0.4, which is comparable to the values found in
experiments. Thus we see that the upper temperature for which the
asymptotic behavior can be observed clearly depends on the quantity
investigated. Note that this observation is in accordance with MCT
since the theory predicts that the magnitude of the corrections to the
asymptotic behavior will depend on the quantity considered.\par

The exponent $\gamma$ of the power-law for $D(T)$ (see Eq.~(\ref{eq3}))
is 2.0 for the A particles and 1.7 for the B particles. Although MCT
predicts these two exponents to be the same, a 10\% deviation from an
asymptotic result is not surprising and therefore not a severe
contradiction to this prediction of the theory. However, in a different
work \cite{kob94a} we have analyzed the temperature dependence of the
$\alpha$-relaxation time $\tau$ (see Eq.(\ref{eq2})) and found, in
accordance with the prediction of MCT, that at low temperatures $\tau$
shows a power-law behavior. MCT predicts that the exponent of the
power-law for $\tau(T)$ and the exponent in the power-law for $D(T)$
should be the same. Since we found that the former is about 2.6
\cite{kob94a} and we now find that the latter is around 1.9 we conclude
that this prediction of the theory is not correct for our system.

Since the connection proposed by MCT between the von Schweidler
exponent $b$ and the critical exponent $\gamma$ \cite{bibles} would
imply $\gamma$=2.8 (using $b=0.49$, which we determined in
Ref.~\cite{kob94a}) we tested whether a plot of $D^{1/\gamma}$ with
$\gamma=2.8$ versus $T$ gives a straight line in some temperature
interval. This would imply that in this temperature interval a
power-law with exponent 2.8 would fit the data well. We found that for
the A particles the data points for $T\leq 0.6$ lie reasonably well on
a straight line. This is not the case for the B particles in any range
of temperatures. Furthermore, the critical temperature that is obtained
for the A particles is around 0.40. This is significantly smaller that
the critical temperatures we determined by other means and which were
all around 0.435 \cite{kob94a,kob94d}. Thus we think that a power-law
with an exponent of 2.8 and a temperature around 0.435 is inconsistent
with our data for the diffusion constant. From the theoretical point of
view it is of course interesting to find that the relaxation times of
the intermediate scattering function for nonzero values of $q$ behave
the way MCT predicts\cite{kob94a,kob94b,kob94d} whereas the diffusion
constants, related to quantities at $q=0$, do not follow these
predictions as closely. To understand this observation it probably will
be necessary to increase our understanding on the corrections of the
asymptotic expressions of the theory for small values of $q$ and we
hope that some progress will be possible in this direction in the
future.

Also included in Fig.~\ref{fig3} is the result of a fit to the
diffusion constants with a Vogel-Fulcher law, i.e. $D\propto \exp
(-B/(T-T_{0}))$. The Vogel-Fulcher temperature $T_{0}$ is 0.268 and
0.289 for the A and B particles, respectively. These two temperatures
are significantly lower than the critical temperatures found for other
quantities, which were all around 0.435 \cite{kob94a,kob94d}. Also, as
can be seen from Fig.~\ref{fig3} in our case the quality of the
Vogel-Fulcher fit is inferior to the one with a power-law. This shows
that our data is good enough to distinguish between the two functional
forms and that therefore the power-law we found is really significant.
Note that this finding is not in contradiction with the situation often
encountered in experiments, where the viscosity, or a relaxation time,
is fitted well by a Vogel-Fulcher law over many orders of magnitude.
The temperatures for which these fits are done are usually closer to
the laboratory glass transition temperature $T_{g}$ than the
temperatures we deal with here. Thus the viscosity is much larger than
the viscosity one would obtain at the lowest temperature investigated
in this work.  Hence our statement is that {\it in the temperature
region investigated here} the diffusion constant is better fitted by a
power-law than by a Vogel-Fulcher law and at present nothing can  be
said about its behavior at lower temperatures.  \par

Since the inverse of the constant of diffusion gives a time scale, we
plotted the MSD versus $tD(T)$. The resulting plot is shown in
Fig.~\ref{fig4}. The curves to the left correspond to low temperatures
and those to  the right to high temperatures. We recognize from this
figure that for intermediate and low temperatures the curves fall onto
a master curve. A comparison with Fig.~\ref{fig2} shows that this
master curve is present for those times that fall into the
$\alpha$-relaxation regime. Thus it can be expected that the master
curve has something to do with the time temperature superposition
principle (see Eq.~(\ref{eq2n})). We thus used as an ansatz a
functional form that is an interpolation between the von Schweidler
behavior at short rescaled times and a diffusive behavior at long
rescaled times, e.g.:
\begin{equation}
\langle r^{2}(t)\rangle=A(r_{c}^{2}+(Dt)^{b})+Dt\quad.
\label{interpol}
\end{equation}
Here $r_{c}, A$ and $b$ are fit parameters. The best fit for the A
particles is included in Fig.~\ref{fig4} as a dashed line. We recognize
that the functional form given by Eq.~(\ref{interpol}) leads to a quite
satisfactory fit in the time region where the master curve is observed.
For the best value of $b$ we obtained 0.48 and 0.43 for the A and B
particles, respectively. These values are in accordance with the value
for the von Schweidler exponent which we found for this system to be
around $b=0.49$~\cite{kob94a,kob94b,kob94d}. Thus MCT is able to
rationalize this master curve quite convincingly.\par

We now turn our attention to a closer examination of the motion of the
particles. This is done conveniently with the help of
$G_{s}^{\alpha}(\mbox{\boldmath $r$},t)$ and
$G_{d}^{\alpha\beta}(\mbox{\boldmath $r$},t)$ ($\alpha,\beta \in
\{A,B\}$), the self and distinct part of the van Hove correlation
function \cite{hansenmcdonald86}. Here and in the following we assume
that the system is isotropic and therefore only the modulus of
$\mbox{\boldmath $r$}$ enters the equations. $G_{s}^{\alpha}(r,t)$ is
defined as
\begin{equation}
G_{s}^{\alpha}(r,t)=\frac{1}{N_{\alpha}}
\left\langle \sum_{i=1}^{N_{\alpha}}\delta
\left(r-|\mbox{\boldmath $r$}_{i}(0)-
\mbox{\boldmath $r$}_{i}(t)|\right) \right\rangle
\stackrel{t,r\rightarrow \infty}{\longrightarrow}
\frac{1}{(4\pi Dt)^{3/2}}\exp\left(-\frac{r^{2}}{4Dt}\right)
\quad ,
\label{eq5}
\end{equation}
where $\delta(r)$ is the $\delta$-function. If $G_s(\mbox{\boldmath
$r$},t)$ depends only on the modulus of $\mbox{\boldmath $r$}$ the
angular integration can be carried out and thus one usually considers not
$G_s(\mbox{\boldmath $r$},t)$ but $4\pi r^{2}G_{s}(r,t)$.\par

The distinct parts $G_d^{\alpha\beta}(r,t)$ are defined by
\begin{equation}
G_{d}^{\alpha\alpha}(r,t)=
\rho g_{\alpha \alpha}(r,t) =
\frac{N_{A}+N_{B}}{N_{\alpha}(N_{\alpha}-1)}
\left\langle\sum_{i=1}^{N_{\alpha}} \sum_{j=1}^{N_{\alpha}}\mbox{ }\!\!'\,
\delta \left(r-|\mbox{\boldmath
$r$}_{i}(0)-\mbox{\boldmath $r$}_{j}(t)|\right) \right\rangle
\stackrel{t,r\rightarrow \infty}{\longrightarrow} 1
\label{eq6}
\end{equation}
and
\begin{equation}
G_{d}^{AB}(r,t)=
\rho g_{AB}(r,t) =
\frac{N_{A}+N_{B}}{N_{A}N_{B}}
\left\langle\sum_{i=1}^{N_{A}} \sum_{j=1}^{N_{B}}
\delta \left(r-|\mbox{\boldmath
$r$}_{i}(0)-\mbox{\boldmath $r$}_{j}(t)|\right) \right\rangle
\stackrel{t,r\rightarrow \infty}{\longrightarrow} 1
\quad .
\label{eq7}
\end{equation}

Here $\rho$ is the density of the system. The prime in the second sum
of Eq.~(\ref{eq6}) means that the term $i=j$ has to be left out.  Note
that we define $G_{d}^{\alpha\alpha}(r,t)$ in a slightly different way
than it is usually done in the literature (see e.g.
\cite{hansenmcdonald86}) in that we divide on the right hand of
Eq.~(\ref{eq6}) by $N_{\alpha}(N_{\alpha}-1)$ instead of the usual
$N_{\alpha}^{2}$.  For a finite system this choice makes
$g_{\alpha\alpha}(r,t)$ approach one for $r$ and $t$ large instead of
$1-1/N_{\alpha}$ as it is the case with the usual definition.\par

In Fig.~\ref{fig5} we plot $4\pi r^{2} G_{s}(r,t)$ for the A particles
for a high (a), an intermediate (b) and a low temperature (c). The
function is presented only for $0\leq r \leq 0.6$ since this is the
most relevant region at low temperatures. In each panel the topmost
curve is for $t\approx 0.32$. The lower curves are then each a factor
of approximately two in time apart. From this figure we recognize that
at high temperatures the self part of the van Hove correlation function
decays in a regular way, e.g.  as in a normal liquid.  This changes a
bit for intermediate temperatures.  Here we see that for times between
$5 \leq t \leq 80$ the curves show a weak tendency to cluster for
$0.05\leq r \leq 0.25$. This effect is much more pronounced at the
lowest temperature investigated, where it can be observed for times
between $2.4 \leq t \leq 640$, which belong to the
$\beta$-relaxation regime\cite{kob94a,kob94d}. This clustering is the
signature that the movement of the particles has dramatically slowed
down in this time interval.  We note that this slowing down takes place
for small distances, thus the particle has not yet left the cage formed
by the particles that surrounded it at time zero. The particle is
typically able to leave this cage, i.e. to move for net distances of
the order of unity, only on a much longer time scale. Thus the process
that takes place in the time window corresponding to the late
$\beta$-relaxation is related to the breaking up of this cage.  MCT
predicts that in this time window the correlation functions should show
a power-law behavior in time (see e.g. Eq.~(\ref{eq2})). That this is
indeed the case is demonstrated in
references\cite{kob94a,kob94b,kob94d}. Note that these distances are
comparable to the one that we found for the height of the plateau in
the MSD (see Fig.~\ref{fig2}) and that also the time window in which we
observe the clustering of the correlation functions is the same as the
one in which we observed the plateau in the MSD. Thus this is a
confirmation that the plateau in the MSD is indeed a signature of the
$\beta$-relaxation process. \par

The idealized version of MCT is based on the assumption that the
so-called ``hopping processes'' are not important contributors to the
relaxation.  This version of the theory makes a number of
straightforward and easily testable predictions that can be compared
with experiments or computer simulations. The more general version of
the theory that takes hopping processes into account is more difficult
to test and thus this has been done only in a few occasions
\cite{cummins93,du94,baschfu94}. We now present evidence that for our
system such hopping processes are not important in the temperature
range we have studied.  This justifies the use of
the idealized theory to interpret the data.

In Fig.~\ref{fig6}a we show $4\pi r^{2} G_s(r,t)$ for the A particles
for times $1.16\leq t \leq 100000$ at $T=0.466$, the lowest
temperatures investigated. The times corresponding to the different
curves were chosen in such a way that the ratio of the times belonging
to two consecutive curves is approximately constant. Thus these times
are, on a {\em logarithmic} time axis, equidistant. Analogous pictures
for a binary soft sphere system or a binary Lennard-Jones system show
that for intermediate times, i.e. a few hundred time units, the
correlation function shows a small peak at distances around one (see,
e.g., Fig.~3 in Ref.~\cite{roux89} or Fig.~4 in
Ref.~\cite{wahnstrom91}). The existence of this peak was interpreted as
an activated process in which a particle hops to the position formerly
occupied by one of the particles in the cage that surrounded it at
$t=0$ \cite{roux89}.  (In this process, the latter particle, or perhaps
some other particle forming the initial cage, presumably hops into the
center of the cage.) These types of processes are not included in the
idealized version of MCT but are approximately accounted for in the
extended version of the theory. A close inspection of Fig.~\ref{fig6}a
shows that for the A particles there is no hint of the presence of a
small secondary peak at distances around one (the location of the
nearest neighbor shell). Thus we can conclude that for the A particles
hopping processes in the above mentioned sense are not present at all.
For the B particles the situation is a bit different.  In
Fig.~\ref{fig6}b we show  $4\pi r^{2} G_s(r,t)$ for the B particles for
the same time interval as in Fig.~\ref{fig6}a. We see that for times
around 15000 time units we observe a secondary peak located around
$r=1.0$. Thus we can conclude that the B particles show some kind of
jump motion. The reason for this difference in the dynamics of the A
and B particles is probably the difference in size of the two types of
particles.  Since the B particles are smaller than the A particles,
they are more mobile, as can also be recognized from the fact that the
diffusion constant for the B particles is larger than the one for the A
particles, and that they therefore are able to make some movements
(jumps) that the A particles cannot do. However, despite the fact that
the B particles show the occurrence of hopping processes we also
recognize from the figure that no secondary peak is observed for times
less than $10^{4}$, i.e.  that these processes are effective only on
the time scale of $10^{4}$ time units or more. For the temperatures
investigated in this work this time scale belongs to the time scale of
the late $\alpha$-process. Therefore the $\beta$-process is not
affected by these hopping processes at all and hence it is reasonable
to analyze all the data that belongs to the $\beta$-relaxation region
with the {\em idealized} version of MCT. We will come to a similar
conclusion when we will study the distinct part of the van Hove
correlation function.\par

In Fig.~\ref{fig5}c we see that in the $\beta$-relaxation region the
various correlation functions for different times seem to cross all in
one common point, namely at $r\approx0.21$. MCT can rationalize such a
behavior by means of the factorization property stated formally in
Eq.~(\ref{eq1}). If we apply this equation to the correlation function
$\phi(r,t)=4\pi r^{2} G_s(r,t)$ we obtain
\begin{equation}
\phi(r,t)=F(r)+H(r)G(t)\quad .
\label{eq8}
\end{equation}
Here $F(r)$ and $H(r)$ are the $r$-dependent offset $f^{c}$ and
amplitude $h$, respectively, from Eq.~(\ref{eq1}). Thus if $H(r)$
becomes zero for some value of $r$ the whole time dependence of the
right hand side disappears and for this particular value of $r$ the
correlation functions are independent of time (note that this statement
is true only in the $\beta$-relaxation region).  This is exactly what
happens around $r\approx 0.21$.\par

In order to make a more stringent test on whether the factorization
property, as stated in Eq.~(\ref{eq8}), really holds we applied a test
which was proposed by Signiorini {\it et al} \cite{signorinijlbmlk90}.
If $t$ and $t'$ are two different times in the $\beta$-relaxation
regime it is easy to show that if the factorization property holds
we have
\begin{equation}
\frac{\phi(r,t)-\phi(r,t')}{\phi(r',t)-\phi(r',t')}=
\frac{H(r)}{H(r')}\quad ,
\label{eq9}
\end{equation}
where $r'$ can be chosen arbitrary. Hence the left hand side of
Eq.~(\ref{eq9}) is independent of $t$ and depends only on $r$.
Signorini {\it et al} showed in their simulation of a molten salt that
if $t'$ is 180ps the left hand is indeed independent of $t$ in
the range $20\mbox{ps} \leq t \leq 40 \mbox{ps}$ and thus the
factorization property holds over this time span.\par

In Fig.~\ref{fig8}a we show the left hand side of Eq.~(\ref{eq9}) for
the A particles in the range $0\leq r \leq 1.5$ at the lowest
temperature investigated, i.e. for $T=0.466$. For $r'$ and $t'$ we
chose 0.13 and 3000 time units, respectively. The time range covered is
$0.02 \leq t \leq 100000$ and for clarity we show only every second
curve in time (as in Fig.~\ref{fig6} the times for which we plot
the curves are spaced evenly on a {\em logarithmic} time axis).
{}From the figure we recognize that if $t$ varies over this time range
(i.e. from microscopic times up to times at the end of the
$\alpha$-relaxation) the left hand side of Eq.~(\ref{eq9}) shows a
significant dependence on time.  To show this a bit clearer we plotted
the curves belonging to the time range before the $\beta$-relaxation as
dotted lines, the curves belonging to the $\beta$-relaxation as solid
lines and the curves for times after the $\beta$-relaxation as dashed
lines.  To see whether the factorization property holds for times in
the $\beta$-relaxation regime we plot in Fig.~\ref{fig8}b only those
curves that correspond to times that are in this regime, thus $3\leq
t \leq 1868$. In order not to overcrowd the figure, only every third
curve in time is shown. We clearly see that for this time range the
curves show only a weak dependence on time and that therefore the
factorization property holds.  The master curve we find is quite
similar to the one predicted by MCT for a binary mixture of soft
spheres\cite{fuchsdiss93} or for hard spheres\cite{fuchs92}. We have to
emphasize that MCT does not predict that $H(r)/H(r')$ is a universal
function. In general it will depend on the system under investigation
and also on the type of particle considered. That $H(r)/H(r')$ is
actually dependent on the type of particle is shown in Fig.~\ref{fig8}c
where we show the same quantity as in Fig.~\ref{fig8}b but this time
for the B particles, for $r'=0.15$ and the time interval $3 \leq t \leq
2113$.  Again we find a master curve if the time lies in the
$\beta$-relaxation regime.  Although the general shape of this master
curve is similar to the one we found in the case for the A particles
there are quantitative differences. For example we recognize that the
two master curves becomes zero at a different value of $r$ thus showing
that the master functions are not universal. Finally we mention that we
found that the form of these master curves for $H(r)/H(r')$ depend only
weakly on the particular choice of $r'$ and $t'$, an observation that
is also in accordance with MCT.\par

A different way to analyze the behavior of the self part of the van
Hove correlation function is by means of the non-gaussian parameters
$\alpha_{n}(t),\quad n=2,3,\ldots$\cite{rahman64boonyip}. These
parameters are measures of the deviation of a distribution function
from a Gaussian form, and it has been proposed that the first of these
parameters, i.e.  $\alpha_{2}(t)$, can be used as an order parameter
for the glass transition occurring at $T_{g-sim}$\cite{odagakiyh91}.
The definition of $\alpha_{2}(t)$ is given by
\begin{equation}
\alpha_{2}(t)=\frac{3 \langle r^{4}(t) \rangle}
{5\langle r^{2}(t) \rangle ^{2}}-1 \quad .
\label{eq10}
\end{equation}
In Fig.~\ref{fig9} we show $\alpha_{2}(t)$ as a function of $t$ for the
A and B particles at intermediate and low temperatures. We see that at
these temperatures the self part of the van Hove correlation function
clearly shows a strongly non-gaussian behavior. This effect becomes
more and more pronounced the lower the temperature is. Also, a
comparison with Fig.~\ref{fig1} shows that the maximum of
$\alpha_{2}(t)$ occurs around the end of the $\beta$-relaxation region.
Furthermore we observe that in the interval that starts at $t\approx
1$ and ends just before the curves attain their maximum the individual
curves seem to follow a master curve.  Although this master curve is
not that well defined for the A particles (Fig.~\ref{fig9}a) it is
quite well defined for the B particles (Fig.~\ref{fig9}b) and should
therefore be considered as a real effect and not just some sort of
coincidence. Note that no scaling or fitting is involved to generate
these master curves. We have tried to determine of what functional form
these master curves are but due to the noise in the data we were not
able to come to a convincing conclusion. \par

Since the interval in which the master curves are observed coincides
well with the interval in which we observe the $\beta$-relaxation
behavior (see e.g.~Fig.~\ref{fig1}) we are strongly tempted to think
that these master curves have something to do with the scaling behavior
observed in the $\beta$-relaxation region (see, e.g. Fig.~\ref{fig4} or
reference~\cite{kob94a,kob94b,kob94d}). However, one should note that
all the scaling behavior we have discussed previously occurs only when
the time axis is scaled by a relaxation time that is strongly dependent
on temperature. This is not the case in Fig.~\ref{fig9}, where we
observe the scaling when we plot the curves versus (unscaled) time. So
far it is not clear whether MCT is able to rationalize our finding and
therefore theoretical work in this direction would certainly be most
useful.  Furthermore it would also be worthwhile to check whether the
observed scaling behavior is also observed for different kind of
systems.  Unfortunately all published data we are aware of only show
$\alpha_{2}(t)$ versus time on a {\em linear} scale and not versus the
{\em logarithm} of time. Thus the master curve, if present, is not
observable, since on a linear time scale it is shifted very close to the
origin.  Finally we mention that we observed a very similar behavior
for the non-gaussian parameter $\alpha_{3}(t)$, thus showing that this
phenomena is not just a peculiarity of $\alpha_{2}(t)$.\par

We turn now our attention to the distinct part of the van Hove
correlation function $G_{d}^{\alpha\beta}(r,t)$, defined in
Eq.~(\ref{eq6}) and (\ref{eq7}).  In Fig.~\ref{fig10} we show
$G_{d}^{\alpha\beta}(r,t)/\rho$ at $t=0$, which is the same as the
radial distribution function $g_{\alpha\beta}(r)$, for the AA, the AB
and the BB correlation for all temperatures investigated. For clarity
the individual curves for different temperatures have been displaced
vertically (see figure caption for details). For the AA and AB
correlation we see that, when the temperature is lowered, the first
neighbor peak becomes narrower. In addition the second
neighbor peak starts to split around $T=0.8$, a feature often observed
for supercooled liquids.  Since we have given evidence above that in
this simulation the system does not undergo a glass transition, i.e.
that $T_{g-sim}$ is less than the lowest temperature investigated here,
we can conclude that the occurrence of a split second neighbor
peak  has nothing to do with $T_{g-sim}$, but is just a feature of a
strongly supercooled liquid.\par

Whereas the distinct part of the van Hove correlation function for the
AA and AB correlation are very similar to the one found for a one
component simple liquid the correlation function for the BB correlation
is quite a bit different.  This is probably due to the fact that the B
particles are in the minority in this system and also smaller than the
A particles. Thus their packing structure in a dense environment is
quite different from the one of a simple one-component liquid. We
recognize from Fig.~\ref{fig10}c that when the temperature is lowered
the first neighbor peak is reduced to a small shoulder. This
can be understood by remembering that the attraction between two B
particles is smaller than the one between a AA pair or a AB pair. Thus
at low temperature two B particles will try to stay apart and thus the
first neighbor peak will become smaller. Contrary to this trend
the second neighbor peak becomes much larger and, similar to
the corresponding peak in the AA and AB correlation function, splits at
temperatures around 0.8.\par

We also investigated the time dependence of the distinct part of the
van Hove correlation function.  In Fig.~\ref{fig11} we show the
dependence of $G_{d}^{\alpha\beta}(r,t)/\rho$ for the AA particles for
three different temperatures. For relatively high temperatures
($T=2.0$, Fig.~\ref{fig11}a) the correlation function decays without
showing any particular feature. (As in Fig.~\ref{fig6} the times for
which we plot the curves are spaced evenly on a {\em logarithmic} time
axis.) This changes when we lower the temperature to $T=0.6$
(Fig.~\ref{fig11}b).  Now we observe that for intermediate times the
correlation functions start to cluster.  This effect is even more
pronounced at the lowest temperature ($T=0.466$, Fig.~\ref{fig11}c).
This clustering is again the result of the $\beta$-relaxation in which
the relaxation is severely slowed down. From this figure we also
recognize that in the time interval of the $\beta$-relaxation regime
the correlation hole at $r=0$ is still there. Thus no particle from the
nearest neighbor shell of the particle at $r=0$ (or somewhere else) has
managed to enter this hole.  We find a small peak at $r=0$ only for
times that are appreciably larger than the ones belonging to the
$\beta$-relaxation regime, namely for $t\ge10^4$ time units.  This
observation is further support for our conclusion made above that on
the time scale of the $\beta$-relaxation, which starts at a few time
units and ends at a few thousand time units, the hopping processes are
not important. In Fig.~\ref{fig12}a we show the same type of correlator
as in Fig.~\ref{fig11} and at the same temperature as in
Fig.~\ref{fig11}c but this time for the AB correlation function. We see
that in the $\beta$-relaxation regime the behavior of this correlation
function is very similar to the one for the AA correlation function.
However, even for times that belong to the $\alpha$-relaxation regime
we do not see any sign of a peak at the origin. Thus we have evidence
that for this correlation function hopping processes are not important
even on the time scale of the $\alpha$-relaxation. This doesn't seem to
be the case for the B particles. In Fig.~\ref{fig12}b we show the
function for the BB correlation. For intermediate times we find again
the clustering of the curves and no peak at the origin.  However, for
longer times there is a pronounced peak at small distances. This peak
is definitely larger than the one observed for the AA correlation
(Fig.~\ref{fig11}c). Thus this finding is in accordance with the
conclusion we made above with regard to the self part of the van
Hove correlation function, namely that in the time region of the
$\alpha$-relaxation the hopping processes are more important for the
dynamics of the B particles than for the A particles.\par

We now proceed to test whether the factorization property holds also
for the distinct part of the van Hove correlation function. We do this
in the same way we tested this property for the self part, i.e. by
checking whether the left hand side of Eq.~(\ref{eq9}) is independent
of time. In Fig.~\ref{fig13} we show the result of this kind of
analysis. Fig.~\ref{fig13}a shows the left hand side of Eq.~(\ref{eq9})
for the AA correlation. The value of $r'$ is 1.05 and the one of $t'$
is 3000 time units. The curves are shown for times between $2.7 \leq t
\leq 700$, thus covering almost two and a half decades in the
$\beta$-relaxation regime.  We clearly see that the curves lie on a
master curve. A comparison with the prediction of MCT for the master
curve of a hard sphere system\cite{barratwgal89} shows that the form of
this master curve is qualitatively similar to the one we find. A
similar result was found by Signorini {\it et al} in a simulation of a
molten salt\cite{signorinijlbmlk90}. A qualitatively similar master
curve is obtained for the AB correlation function, which is shown in
Fig.~\ref{fig13}b. Here we chose $r'=0.90$ and $t'=3000$. The time
range covered is $3.1 \leq t \leq 1142$.  Although the master curve is
of similar shape as the one for the AA correlation the details of the
two master curves differ, thus showing again, as in the case of the self
part of the van Hove correlation function, that these master curves are
not universal but depend on the type of correlator investigated.  This
is shown even more clearly with the master curve found for the BB
correlation function (Fig.~\ref{fig13}c).  To compute it we chose
$r'=1.40$ and $t'=3000$. The time range shown is $2.1 \leq t \leq
1142$. We recognize that for this correlation function the shape of the
master curve is very different from the one for the AA and AB
correlation function, and it would be interesting to see whether MCT is
able to rationalize also a master curve like this.

\section{Summary and Conclusions}

We have presented the results of a large scale computer simulation we
performed in order to test the correctness of the predictions of MCT
for a supercooled binary Lennard-Jones liquid.  In this work we
concentrated on the investigation of the mean squared displacement of a
tagged particle and on the van Hove correlation function. The results
can be compared with some of the findings of our investigations of the
intermediate scattering function\cite{kob94a,kob94b,kob94d} in order to
test whether MCT is able to give a correct description of
the dynamics of the system investigated.\par

We have strong evidence that for {\it all} temperatures investigated
these results are all {\it equilibrium} properties of the system and
are therefore independent of cooling-rates or the thermal history of
the system. This evidence includes the observation that the pressure
and the potential energy of the system are smooth functions of
temperature, i.e. show no sign of a singularity of some sort.
Furthermore we show that the mean squared displacement (MSD) of a
tagged particle shows a diffusive behavior at long times. Moreover we
find that all correlation functions investigated decay to zero within
the time span of our simulation.\par

We find that, for low temperatures, the curves of the MSD collapse onto
a master curve when they are plotted versus $tD(T)$, where $D$ is the
constant of diffusion. This master curve is observed in a time range
where we find that also the intermediate scattering function shows a
scaling behavior and which has been identified with the
$\alpha$-relaxation regime\cite{kob94a,kob94b,kob94d}. The master curve
is fitted well by an interpolation formula between the von Schweidler
behavior at short rescaled times and the diffusive behavior at long
rescaled times. The exponent of the von Schweidler law in this
interpolation formula is close to the value we found for the von
Schweidler exponent of the intermediate scattering function of this
system\cite{kob94a,kob94b,kob94d}. This is further evidence that the
scaling behavior in the MSD is a consequence of the scaling behavior
predicted by MCT in the $\alpha$-relaxation region.\par

The diffusion constant of both types of particles show a power-law
behavior at low temperatures. The critical temperature $T_{c}$ is for
both species the same and also the critical exponents $\gamma$ differ
only by about 10\%. Thus this is in accordance with MCT. However, the
value of $\gamma$ does not match the value
of the critical exponent that we found for the divergence of the
relaxation time of the intermediate scattering
function.\cite{kob94a,kob94d}. Thus this result is in
disagreement with MCT.\par

Note that this disagreement is a bit surprising since we have shown in
references \cite{kob94a,kob94d} that various correlators show a scaling
behavior with a scaling time that shows a power-law dependence on
temperature. The critical exponent of this power-law fulfills the
connection predicted by MCT between this exponent and the von
Schweidler exponent $b$, which was extracted from the scaling behavior
of the correlators. Here we find now that the MSD shows a scaling
behavior too. The scaling function seems to be in accordance with the
one found for the correlators. The corresponding time scale, i.e. the
inverse of the diffusion constant, shows a power-law behavior with the
same critical temperature as the relaxation times of the mentioned
correlators. However, the exponent is {\em not} the same and thus the
MCT connection between $b$ and the $\gamma$ extracted from $D(T)$ is
not obeyed.\par

The analysis of the self part of the van Hove correlation function
showed that the $\beta$-relaxation regime can be recognized in this
quantity as well. With the help of this function we also gave evidence
that hopping processes are unimportant for the dynamics on the
time-scale of the $\beta$-relaxation.  This justifies use of the
idealized version of MCT, which neglects such hopping processes, to
interpret the data.  Also on the time scale of the $\alpha$-relaxation
we do not see any indication that these hopping processes are present
for the A particles. However, for the B particles we see that on the
time scale of the {\em late} $\alpha$-relaxation a second
neighbor peak appears and that therefore hopping processes become
relevant. Note that this observation, namely that hopping processes
affect different quantities in different degrees, is not in
contradiction with MCT.\par

We have tested whether the factorization property proposed by MCT, see
Eq.~(\ref{eq9}), holds for this type of correlation function. This test
was performed in a way proposed by Signorini {\it et
al}\cite{signorinijlbmlk90}. We found that in the time region of the
$\beta$-relaxation regime the factorization property actually holds.
Thus we confirmed the prediction of MCT on the existence of this
property and the time range in which it is supposed to hold. Also the
form of the curve of the critical amplitude $H(r)$ is in qualitative
agreement with the prediction of MCT for a binary soft sphere
system\cite{fuchsdiss93}.\par

The investigation of the three parts of the distinct part of the
van Hove correlation function, i.e. the AA, the AB and the BB
correlation function, lead to similar conclusions as in the
case of the self part, thus confirming the existence of the
$\beta$-relaxation regime and the absence of hopping processes on the
time scale of this regime. Also in this case we found that the
factorization property holds in the time range of the
$\beta$-relaxation regime and that the critical amplitude $H(r)$ is
qualitatively similar to the one predicted by MCT for a hard sphere
system\cite{barratwgal89}.\par

In addition to these observations, which can all be rationalized in the
framework of MCT, we have also found that the non-gaussian parameters
$\alpha_{2}(t)$ and $\alpha_{3}(t)$ show a master curve when they are
replotted versus time, i.e. {\em not} versus {\em rescaled} time. So
far it is not clear whether this observation can be understood with
the help of MCT or whether it is in contradiction to the theory.
Therefore it would certainly be important to see whether this feature
is found also in different types of systems.\par

In summary we can say that there are a fair number of features in the
dynamics of the Lennard-Jones mixture investigated that can be
rationalized in the framework of the idealized version of the MCT.  The
idealized version predicts the existence of a dynamical singularity at
a temperature $T_c$ and predicts several signatures of this singularity
in the behavior of the system above $T_c$.  These signatures include,
for example, apparent power law dependence of relaxation rates on
temperature and the factorization property in the $\beta$-relaxation
regime.  The Lennard-Jones mixture studied exhibited many of these
signatures in a temperature range corresponding to a value of
$\epsilon=(T-T_c)/T_c$ between about 0.07 and 0.8. (The singularity is
not actually observed, but its apparent temperature is inferred from
the higher temperature data.) Thus we conclude that the idealized MCT
is able to describe some essential features of relaxation in
supercooled liquids.\par

\acknowledgements
We thank Dr. J. Baschnagel, Dr. M. Fuchs and Prof. W. G\"otze for many
useful discussions and a critical reading of the manuscript.  Part of
this work was supported by National Science Foundation grant
CHE89-18841. We made use of computer resources provided under NSF grant
CHE88-21737.

\begin{figure}
\noindent
\caption{Total energy (solid line), potential energy (dashed line)
and pressure (divided by 10, dotted line) versus temperature $T$.\label{fig1}}
\vspace*{5mm}
\par
\caption{Mean squared displacement versus time for A particles for
all temperatures investigated.\label{fig2}}
\vspace*{5mm}
\par
\caption{Diffusion constant $D$ for A and B particles (lower and
upper curve, respectively) versus $T-T_{c}$. The critical temperature
$T_{c}$ is 0.435. Also shown are the power-law fits with exponents
2.0 and 1.7 for the A and B particles, respectively. The dashed lines
are the best fits with a Vogel-Fulcher law.\label{fig3}}
\vspace*{5mm}
\par
\caption{Mean squared displacement versus $tD(T)$ for the A particles
(solid lines) for all temperatures investigated. The low temperatures
are to the left and the high ones to the right. Dashed curve: Best fit
to the master with the functional form of Eq.~(5).\label{fig4}}
\vspace*{5mm}
\par
\caption{$4\pi r^{2} G_{s}(r,t)$ for the A particles versus $r$ for times
$t\approx 0.32\cdot 2^{n}$ with $n=0,1,2,\dots$. a) $T=2.0$, b) $T=0.6$,
c) $T=0.466$.\label{fig5}}
\vspace*{5mm}
\par
\caption{$4\pi r^{2} G_{s}(r,t)$ for the A particles (a) and the B
particles (b) at $T=0.466$ for times between 1.16 and $10^{5}$ time
units. Note the presence of a small secondary peak for times around
10000 for the B particles. No such peak is observed for the A
particles.\label{fig6}}
\vspace*{5mm}
\par
\caption{Normalized critical amplitude $H(r)/H(r')$ for $0 \leq r \leq
1.5$ (see Eq.~(10) ) for $T=0.466$, $t'=3000$. a) A particles: $0.02
\leq t \leq 100000$ (only every second correlation function is shown),
$r'=0.13$. Dotted lines: $0.02 \leq t < 3$, solid lines: $3\leq t\leq
1868$, dashed lines: $1868 < t \leq 100000$. b) A particles $3 \leq t
\leq 1868$ (only every third correlation function is shown), $r'=0.13$.
c) B particles: $3 \leq t \leq 2113$ (only every third correlation
function is shown), $r'=0.15$.\label{fig8}}
\vspace*{5mm}
\par
\caption{Non-gaussian parameter $\alpha_{2}$ versus $t$ for the A
particles (a) and the B particles (b). Temperatures from right to left:
0.466, 0.475, 0.5, 0.55, 0.6, 0.8, 1.0.\label{fig9}}
\vspace*{5mm}
\par
\caption{Radial distribution function $g(r)$ for AA (a), AB (b) and
BB (c) correlation for all temperatures investigated. For clarity the
individual curves have been shifted vertically by $0.15n,
n=1,2,3\ldots$.\label{fig10}}
\vspace*{5mm}
\par
\caption{$G_{d}(r,t)/\rho$ for the AA correlation function. a)
$T=2.0$, $0\leq t \leq 400$; b) $T=0.6$,$0 \leq t \leq 3000$;
c) $T=0.466$, $0 \leq t \leq 100000$.\label{fig11}}
\vspace*{5mm}
\par
\caption{$G_{d}(r,t)/\rho$ for $T=0.466$ and $0\leq t \leq 100000$. a)
AB correlation, b) BB correlation.\label{fig12}}
\vspace*{5mm}
\par
\caption{Normalized critical amplitude $H(r)/H(r')$ for distinct part
of the van Hove correlation function at $T=0.466$ (see Eq.~(10)).
$t'=3000$. a) AA correlation; $r'= 1.05$, $2.7 \leq t \leq 700$. b) AB
correlation; $r'=0.90$, $3.1 \leq t \leq 1142$. c) BB correlation;
$r'=1.40$, $2.1 \leq t \leq 1142$.\label{fig13}}
\vspace*{5mm}
\par
\end{figure}
\end{document}